\documentclass[fleqn,twoside,11pt]{article}

\usepackage[numbers]{natbib}
\usepackage{left_eq}   
\usepackage{CSPM}  

\usepackage[T1]{fontenc}

\usepackage[latin1]{inputenc}

\usepackage[english,francais]{babel}

\usepackage[cyr]{aeguill}

\usepackage{pstricks, pst-plot}	
\usepackage{graphicx} 
\usepackage{wrapfig}  
\usepackage[figuresright]{rotating}
\usepackage{subcaption}
\usepackage{float}

\usepackage{amsmath}
\usepackage{amssymb}

\usepackage{amsmath,mathrsfs,color,mathdots,bbold,tikz,xspace,version}
\usepackage{amsthm}

\usepackage[english]{babel}
\usepackage{blindtext}

\newcommand\be{\begin{equation}}
\newcommand\ee{\end{equation}}

\newcommand\bea{\begin{eqnarray}}
\newcommand\eea{\end{eqnarray}}

\newcommand{\ud}{d}

\DeclareMathOperator{\tr}{tr}

\newcommand{\ket}[1]{\ensuremath{|#1\rangle}\xspace}
\newcommand{\bra}[1]{\ensuremath{\langle #1|}\xspace}
\newcommand{\moy}[1]{\ensuremath{\langle #1 \rangle}\xspace}
\newcommand{\tra}[1]{\ensuremath{\tr \left\{ #1 \right\}}\xspace}

\newcommand{\dt}[1]{\ensuremath{ \frac{{d} #1}{{d} t} }\xspace}

\def\R{\mathbb R}

\def\C{\mathbb C}

\def\Hs{{\mathfrak H}}

\def\Bs{{\mathfrak B}}

\def\id{{\mathbb 1}}

\def\|{{|\:\!\!|}}

\def\HS{{\mathfrak H}_s}

\def\Li{{\cal L}}

\title{Open extended quantum systems}
\author{Dragi Karevski}
\affil{Universit\'e de Lorraine, CNRS, LPCT, F-54000 Nancy, France}
\email{dragi.karevski@univ-lorraine.fr}

 \HeaderTitle{Open extended quantum systems}
 \HeaderAuthor{D. Karevski}

\begin{document}
\maketitle             

\thispagestyle{empty}  

\begin{changemargin}{1.5cm}{1.5cm} 
\begin{abstract}
  \small{We present an introduction to the theory of open extended quantum systems.
We begin with a microscopic derivation of the so-called Lindblad equation followed by a more abstract approach. 
Next, we introduce collision models, a versatile framework that offers a possible unraveling of the non-unitary dynamics of open quantum systems.  We finally discuss
concrete situations involving quantum transport phenomena, the generation and replication of entanglement or even the non-thermal relaxation of cold atomic gases confined in optical traps. 
} 
\end{abstract}

\end{changemargin} 


\section{Open quantum dynamics}
A physical system is never completely isolated. Inevitable residual interactions with the environment will affect the system's unitary dynamics and most often lead to dissipation and  loss of quantum coherences.  
However, if one believes that quantum mechanics must be valid at all scales, the dynamics of the system plus its environment, taken as a whole, must obey the axioms of quantum theory and therefore be unitary within the total Hilbert space $\Hs= \Hs_S \otimes \Hs_E$, where $\Hs_S$ and $\Hs_E$ represent the Hilbert spaces of the system and environment, respectively, see \cite{AlickiLendi, Alicki, Breuer, Schaller, Weiss, Karevskiellipse,Attal} for  monographs on the subject. In the following,  for the sake of simplicity we will consider finite dimensional Hilbert spaces only.


The unitary dynamics of the total system is generated by $U(t)=e^{- \frac{i}{\hslash} H_{tot} t }$ with the total Hamiltonian 
$H_{tot} = H^0_{tot}+ V = H_S + H_E + V$
where $H_S=H_s \otimes \id_e$ and $H_E=\id_s \otimes H_e$ are the Hamiltonians of the system and the environment and where 
\be
V=  \sum_i X^i \otimes Y^i
\ee
is a sum of product of operators $X^i$ and $Y^i$, associated with the system and the environment, respectively,  and where each product describes a different dissipation channel.  
Without loss of generality, we will assume that the operators $X^i$ and $Y^i$ are hermitian. 

The dissipative channels can either act on the entire system or target specific locations within the extended system. In the latter scenario--such as a linear chain coupled at different sites to distinct reservoirs--a current is expected to develop within the system, eventually leading, at sufficiently long times, to a non-equilibrium steady state \cite{Karevski2009,Platini2010,Prosen2011,Karevski2013,Popkov2013,Landi2014,Landi2015}.  If all reservoirs share identical properties, the system is expected to relax to a currentless state, which may or may not correspond to a Gibbs state \cite{Dhahri2008,Attal2014}.

\subsection{Microscopic theory: weak coupling limit}
In the weak coupling limit, it is assumed that the interaction term  is much smaller than $H_s$: $\|V\|={\cal O}(\epsilon\|H_s\|)$, and therefore 
the simplest way to proceed is to start with the Liouville equation for the total system in the interaction picture. To do this, let us define the density operator in the interaction picture by the unitary transformation $\rho_I(t)\equiv e^{ \frac{i}{\hslash} H_{tot}^0 t} \rho(t)\; e^{ -\frac{i}{\hslash} H_{tot}^0 t}$. 
The Liouville equation satisfied by $\rho_I(t)$  is given by
\be
\dt{} \rho_I(t)= \frac{1}{i\hslash} [V_I(t), \rho_I(t)]\; ,
\label{liouville1}
\ee
where $V_I(t)\equiv e^{ \frac{i}{\hslash} H_{tot}^0 t}  V  e^{- \frac{i}{\hslash} H_{tot}^0 t}$, with the equivalent formal solution  
\be
\rho_I(t) = \rho_I(0) + \frac{1}{i\hslash} \int_0^t \ud t'\; [V_I(t'), \rho_I(t')]\; .
\label{liouville1int}
\ee
In the following we will assume that the initial condition 
 is given by a tensor product of the initial states of the system and the environment: $\rho_I(0)=\rho(0) =  \rho_s \otimes \rho_e$.
Substituting the formal solution (\ref{liouville1int}) back into (\ref{liouville1}), we obtain
\be
\dt{} \rho_I(t)= \frac{1}{i\hslash} [V_I(t), \rho(0)] - \frac{1}{\hslash^2} \int_0^t \ud t'\; [V_I(t),[V_I(t'), \rho_I(t')]] \; .
\label{liouville1int2}
\ee
In the weak coupling limit, since it is assumed that $V\propto \epsilon$, this expression shows that the double commutator generates a term which is at least of the order $\epsilon^2$.  
Noticing that $\tr_E\{\rho_I(t)\} = e^{\frac{i}{\hslash} H_s t} \rho^s(t) e^{-\frac{i}{\hslash} H_s t} \equiv \rho^s_I(t)$,
one obtains form the previous equation, by taking the trace over the environment, the system's dynamical equation  
\be
\dt{} \rho^{s}_I(t)= \frac{1}{i\hslash}  \tr_E\{[V_I(t),\rho(0)] \}  - \frac{1}{\hslash^2} \int_0^t \ud t'\; \tr_E\{ [V_I(t),[V_I(t'), \rho_I(t')]] \} \; .
\ee
Since the initial state $\rho(0)=\rho_s \otimes \rho_e$ is a product state, the first term in the right hand side of this equation is simply given by 
$\sum_i [X^i_I(t), \rho^s] \moy{Y^i}^0(t)$,
where $\moy{Y^i}^0(t)= \tr_E \{Y_I^i(t) \rho_e\}$. If we assume that the state of the environment $\rho_e$ is stationary under the free dynamics generated by $H_e$, which is the case for a typical Gibbs state, then the average $\moy{Y^i}^0(t)$ is time-independent and one can always cancel these terms by a proper redefinition of the variables $Y^i \rightarrow Y^i - \moy{Y^i}^0$. 
To further simply the problem, we will assume that the environment remains in a steady state, that is $\rho(t')\simeq \rho^s(t')\otimes \rho_e$ and replace under the integral $\rho^s(t')$ by $\rho^s(t)$. This is a Markovian approximation valid at order $\epsilon^2$. 
As a consequence, the time evolution of the system is governed by the equation 
\begin{align}
\dt{} \rho^s_I(t)& =  - \frac{1}{\hslash^2} \int_0^t \ud t'\; \tr_E\{ [V_I(t),[V_I(t-t'), \rho^s_I(t)\otimes \rho_e]] \} \; . 
\label{liouville1int5}
\end{align}

\subsubsection{Spectral decomposition} 
In order to proceed further, one has to take explicitly the trace over the environment.
To do so, let us use the eigen-basis $\{\ket{n}\}$ of $H_s$ with associated eigen-energies $\{\epsilon_n\}$ and write 
\be
X^i = \sum_n \sum_m \bra{n} X^i \ket{m} \ket{n} \bra{m} = \sum_\omega \tilde{X}^i (\omega)
\ee
where the last sum is taken over the Bohr frequencies  $\hslash \omega =\epsilon_m -\epsilon_n$ with 
\be
\tilde{X}^i (\omega)  
=  \sum_{\epsilon_m-\epsilon_n=\hslash \omega} \langle n |X^i|m\rangle |n\rangle \langle m |\; . 
\ee
Noticing that  $[H_s, \tilde{X}^i (\omega )] =-\hslash \omega  \tilde{X}^i (\omega )$, we see that the spectral operators $\tilde{X}^i (\omega) $ are jump operators between energy levels spaced by $\hslash \omega$. 
The time-dependence of $X^i$ is thus simply expressed as 
\be
X^i(t) =\sum_\omega e^{-i\omega t} \tilde{X}^i (\omega) \; . 
\label{decompXomega}
\ee
Injecting (\ref{decompXomega}) into (\ref{liouville1int5}) one obtains for the term proportional to  $\tr_E\{V_I(t) V_I(t-t') \rho^s_I(t)\otimes \rho_e \}$
\begin{align}
 - \frac{1}{\hslash^2} \sum_{ij}   \sum_{\omega\omega'}   e^{i(\omega -\omega')t }
  { \tilde{X}^i (\omega)}^\dagger  \tilde{X}^j (\omega' ) \rho^s_I(t)     \int_0^t \ud t'\;   e^{i\omega' t'  }  \moy{Y^i(t') Y^j}^0
\end{align}
Assuming that the correlations in the environment decay sufficiently rapidly, we can extend the upper bound of the integral  up to infinity. Introducing the environment spectral functions
\be
\Gamma^{ij}(\omega) \equiv \int_0^\infty \ud t\;   e^{i\omega t  }  \moy{Y^i(t) Y^j}^0\; ,
\ee
the previous term can be written in the Markov approximation as
\be
- \frac{1}{\hslash^2} \sum_{ij}   \sum_{\omega\omega'}   e^{i(\omega -\omega')t } \Gamma^{ij}(\omega')
  { \tilde{X}^i (\omega)}^\dagger  \tilde{X}^j (\omega' ) \rho^s_I(t) \; . 
\ee
Performing a secular approximation, that is, by assuming that the terms $e^{i(\omega -\omega')t } $ oscillate very rapidly, we can select only the diagonal terms $\omega=\omega'$ in the double sum. Gathering all the terms coming from the double commutator in (\ref{liouville1int5}) one arrives at
\begin{align}
\dt{} \rho^s_I(t) &=   \sum_{ij}   \sum_{\omega}   \frac{ \Gamma^{ij}(\omega)  }{\hslash^2} 
 \left(   \tilde{X}^j (\omega) \rho^s_I(t)   { \tilde{X}^i (\omega)}^\dagger  -{ \tilde{X}^i (\omega)}^\dagger  \tilde{X}^j (\omega ) \rho^s_I(t) 
  \right)    + \; \textrm{h. c.}  \; . 
\label{liouville1int7}
\end{align}
Decomposing the matrix $\Gamma(\omega)$ into $\Gamma (\omega ) \equiv \frac{1}{2} \left( \gamma(\omega) + i\sigma(\omega)  \right)$, 
where $\gamma$ and $\sigma$ are Hermitian matrices,  after small rearrangements the previous equation becomes 
\begin{align}
\dt{} \rho^s_I(t) = -\frac{i}{\hslash} \left[\Delta_l, \rho^s_I(t)\right] 
+ \sum_{i,j,\omega} \frac{\gamma^{ij}(\omega)}{\hslash^2} \left(  \tilde{X}^j (\omega) \rho^s_I(t)   { \tilde{X}^i (\omega)}^\dagger   
 -   \frac{1}{2}  \left\{ { \tilde{X}^i (\omega)}^\dagger  \tilde{X}^j (\omega ), \rho^s_I(t) \right\} \right) \; ,
\end{align}
where we recognize in the commutator term a so-called Lamb shift  $\Delta_l \equiv \sum_{ij\omega} \hslash \sigma^{ij} (\omega) { \tilde{X}^i (\omega)}^\dagger  \tilde{X}^j (\omega )$
generating an additional unitary contribution which commutes with the system Hamiltonian $H_s$.

Going back to the Schrödinger picture, one restores the unitary contribution generated by the system Hamiltonian $H_s$ and we finally get the so-called Lindblad equation or more properly dubbed the Lindblad-Gorini-Kossakowski-Sudarshan equation \cite{Lindblad,Gorini}
\begin{align}
\dt{} \rho^s(t) = -\frac{i}{\hslash} \left[H_s + \Delta_l, \rho^s(t) \right]   
+ \sum_{i,j,\omega} \frac{\gamma^{ij}(\omega)}{\hslash^2} \left(  \tilde{X}^j (\omega) \rho^s(t)   { \tilde{X}^i (\omega)}^\dagger  
-  \frac{1}{2}  \left\{ { \tilde{X}^i (\omega)}^\dagger  \tilde{X}^j (\omega ), \rho^s(t) \right\} \right) \; . 
\label{liouville1int8}
\end{align}

The dissipation matrix $\gamma$ being  Hermitian and positive semi-definite, it can be diagonalized by a unitary transformation:
$\gamma V = V  \lambda  \quad  \Leftrightarrow \quad  V^\dagger \gamma V = \lambda$, 
where $\lambda$ is the diagonal matrix containing the eigenvalues  $\lambda_k\ge 0$ of $\gamma$, and $V$  is the matrix of orthonormal eigenvectors $v_k$, i.e., such that  $V_{ik}= v_k(i)$. 
By introducing the diagonal operators $L^k(\omega)   =  \sqrt{\lambda_k(\omega)} \sum_i v_k^*(i,\omega)  \tilde{X}^i(\omega)/\hslash$
for each mode $\omega$, the  Lindblad equation  (\ref{liouville1int8}) takes the standard form
\be
\dt{} \rho^s(t) = -\frac{i}{\hslash} \left[H_s + \Delta_l, \rho^s(t) \right]    + \sum_{\omega ,k}  \left( L^k (\omega) \rho^s(t)   { L^k(\omega)}^\dagger 
- \frac{1}{2}  \left\{ { L^k (\omega)}^\dagger  L^k (\omega ), \rho^s(t) \right\} \right) \; . 
\ee

\subsubsection{Steady state} 
If the environment is in a Gibbs state  $\rho_e \propto e^{-\beta H_e}$ at inverse temperature $\beta$, then the Lindblad equation (\ref{liouville1int8}) admits as a unique solution the Gibbs state $\rho^s_\beta = \frac{e^{-\beta H_s}}{\tr_s\{ e^{-\beta H_s}\}}  $
given that the dynamics is ergodic. Ergodicity of the dynamics is guaranteed by the property \cite{Evans1977,Nigro2019}
\be
[Q, \tilde{X}^i(\omega)] = [Q, {\tilde{X}^i(\omega)}^\dagger]= 0 \quad \forall i,\omega\quad  \Rightarrow \quad  Q\propto \id\; . 
\ee
This condition together with $\rho_e \propto e^{-\beta H_e}$  implie that the system relaxes toward the Gibbs state   $\rho^s_\beta$ whatever the initial state:  
$\lim_{t\rightarrow \infty} \rho^s(t) \rightarrow \rho^s_\beta$.
The stationarity of the Gibbs state $\rho^s_\beta$, thanks to $[H_s,\Delta_l]=0$ and $\rho^s_\beta \tilde{X}^i(\omega) = e^{\beta \omega} \tilde{X}^i(\omega) \rho^s_\beta$,  leads from  the Lindblad equation to the condition 
\be
\sum_{i,j,\omega} \left( \gamma^{ij}(\omega) e^{-\beta \omega} - \gamma^{ji}(-\omega) \right) \tilde{X}^j(\omega) {\tilde{X}^i(\omega) }^\dagger \rho^s_\beta = 0\; ,
\ee
which is obviously satisfied by the detailed balance condition $ \gamma^{ij}(\omega) e^{-\beta \omega} = \gamma^{ji}(-\omega)$.
Detailed balance is induced by the Kubo--Martin--Schwinger condition on the environment correlation functions  $\moy{Y^i(t) Y^j}_\beta = \moy{Y^j  Y^i(t+i\beta)}_\beta$  together with the limit $\lim_{t\rightarrow \infty}    \moy{Y^i(t) Y^j}_\beta = 0$.

One of the particularities of the dynamical equation (\ref{liouville1int8}) is that, in the case of a non-degenerate spectrum of $H_s$, it decouples the dynamics of the populations -- defined by the diagonal elements  $p_n \equiv  \langle n | \rho^s | n \rangle $  of the density matrix in the eigenbasis of the Hamiltonian $H_s$ -- from the dynamics of the off-diagonal terms, the coherences,  $\rho_{nm}\equiv \langle n |\rho^s | m\rangle $.  
Indeed, by taking the expectation on the eigenstate $\ket{n}$ of the Lindblad equation (\ref{liouville1int8}), one obtains the stochastic Master Pauli equation 
\be
\dt{} p_n(t) = \sum_k W_{k\rightarrow n } p_k(t) - W_{n\rightarrow k } p_n(t)  \equiv \sum_k M_{nk} p_k(t)\; , 
\label{eqmaitressePauli}
\ee
where the stochastic matrix  $ M$ is defined for $k\neq k'$ by $M_{kk'} = W_{k'\rightarrow k }$ and $M_{kk} = - \sum_{k'} W_{k\rightarrow k' }$
with the transition rates
\be
W_{k\rightarrow n } =  \sum_{i,j}  \frac{ \gamma^{ij}(\epsilon_{k}-\epsilon_n)  }{\hslash^2}  \langle k|  {X}^i | n \rangle \langle n |  {X}^j  |k\rangle    \; . 
\ee
The fact that the sum of the elements of each column of the stochastic matrix is zero, $\sum_n M_{nk}=0$ garanties the conservation of probabilities  
$\dt{} \sum_n p_n(t)= 0$. The detailed balance condition on the $\gamma^{ij}$s  transfers into 
a detailed balance condition on the transition rates  $e^{-\beta \epsilon_k}  W_{k\rightarrow n } = e^{-\beta \epsilon_n}  W_{n\rightarrow k }$ which in turn implies that the steady state distribution of the populations is the Boltzmann--Gibbs distribution $ \frac{e^{-\beta \epsilon_n}}{\sum_k e^{-\beta \epsilon_k}}$.

\section{Abstract derivation: Completely Positive Dynamics}
In the general case, a Lindblad equation, which therefore does not necessarily arise from a weak coupling limit, can always be put into the diagonal form \cite{Lindblad}
\begin{align}
\dt{} \rho(t) &=  -\frac{i}{\hslash} [H, \rho(t)] + \sum_j \left[ L_j \rho(t) L_j^\dagger - \frac{1}{2} \{ L_j^\dagger L_j, \rho(t)\}\right] \equiv {\cal L}(\rho(t)) 
\label{lindblad2}
\end{align}
where $L_j$, not necessarily Hermitian, are the generators of the Lindblad dynamics and $\rho(t)$ the density operator of the open system--we drop the subscripts $s$, since all the operators are elements of the system algebra. We can also note that $H$ represents the generator of the unitary part of the dynamics and we saw earlier that it is not necessarily the Hamiltonian of the system since it can be shifted by a Lamb contribution. 
The representation of Lindblad dynamics is not unique. Indeed, the simultaneous transformation of the generators $L_i$ and the Hamiltonian $H$ into a 
\be
\left\{\begin{array}{l}
L_j \rightarrow L_j'= L_j + a_j\id\\
H\rightarrow H'=H + \frac{\hslash}{2i} \sum_j (a_i^*L_j -a_jL_j^\dagger) + b\id
\end{array} \right. \; , 
\ee
where $a_j \in \C$ and $b\in \R$, leaves the dynamics invariant: 
$
\dt{}\rho(t) = {\cal L}(\rho(t)) = {\cal L'}(\rho(t))
$.

\subsection{Kraus map}
The Lindblad dynamics is a special case of a completely positive dynamics, the particularity being linked to the markovianity of the dynamics. In the general case, an admissible dynamics is a one-parameter family of transformations $\Lambda_t$, completely positive, which can be put in Kraus form \index{Kraus (decomposition)}
\be
\rho(t) = \Lambda_t \rho_0 = \sum_\alpha K^\alpha \rho_0 { K^\alpha}^\dagger 
\label{Kraus}
\ee
with the condition 
\be
\sum_\alpha{ K^\alpha}^\dagger K^\alpha = \id
\label{Kraus_unity}
\ee
on the generators $K^\alpha \in \Bs(\HS)$. 
This last condition implies the conservation of probabilities over time. Note that a unitary dynamics is also of Kraus form with a single Kraus operator satisfying $K^\dagger K=\id$.

One may introduce the dual dynamics of $\Lambda_t$, i.e. a Heisenberg representation for the super-operators acting on the elements of the Banach space $\Bs(\HS)$, via the usual definition of the adjoint of a linear operator
\be
(A, \Lambda_t B) = (\Lambda_t^{\star} A, B)
\ee 
where the scalar product on $\Bs(\HS)$ is defined by 
\be
(A,B) = \tra{A^\dagger B}\; . 
\ee 
This leads to the identification 
\be
\Lambda_t^{\star} A = \sum_\alpha  {K^\alpha}^\dagger  A  K^\alpha \; . 
\ee
We thus see that the condition (\ref{Kraus_unity}) ensures that the dual dynamics preserves the identity $\Lambda^{\star} \id = \id$, which is another way of expressing the preservation of probabilities: $ (\Lambda_t^{\star} \id, \rho_0) = ( \id, \rho_0)  = 1$.

\subsection{Markovian dynamical map}

A Markovian quantum dynamics is a one-parameter family of dynamical transformations satisfying the following properties:
\begin{itemize}
\item $ \Lambda_t $ is a dynamics ($\Lambda^\star_t$ completely positive and $\Lambda^\star_t \id =\id$).
\item $\Lambda_t \Lambda_s = \Lambda_{t+s}$ semi-group condition or Markov property
\item $\tra{ A \Lambda_t \rho}$ is a continuous function of $t$ for all $A\in \Bs(\HS)$ and all density operator $\rho$. 
\end{itemize}
Given these conditions, we can show that there exists an infinitesimal generator $\Li$ of the dynamics such that  $\Lambda_t = e^{t\Li}$. 
Indeed, in the vicinity of the identity ($t=\epsilon\rightarrow 0$), 
$\Lambda_\epsilon = \id + \epsilon \Li$, 
which gives by the semi-group property $\Lambda_{n\epsilon} = \left( \id + \epsilon \Li\right)^n$. By choosing $\epsilon= t/n$ and taking the limit $n \rightarrow \infty$ we have from the exponential formula
\be
\Lambda_t = \lim_{n\rightarrow \infty} \left( \id + \frac{t}{n}\Li\right)^n = e^{t\Li} \; . 
\ee
By taking the derivative of the equation $\rho(t) = \Lambda_t \rho_0$ we arrive at the master equation
\be
\dt{}\rho(t) = \Li \rho(t) \; . 
\ee 
The most general form of the infinitesimal generator of Markov dynamics defined on a finite-dimensional Hilbert space is Lindblad's and is extracted from the Kraus decomposition \eqref{Kraus}. 

The Kraus generators are elements of $\Bs(\HS)$ and can therefore be decomposed onto a linear basis of operators 
$\{F_k\} \in\Bs(\HS) $ with $k=0,1,\dots , N^2-1$ which can always be chosen such that $F_0=\id$ and the others of zero trace, $\tra{F_{k\ne 0}}=0$. 
With the expansion $K^\alpha = \sum_{k=0}^{N^2-1} a^\alpha_{k}(t) F_k$
the dynamics is written as 
\begin{align}
\Lambda_t \rho_0 &= \sum_{k,l=0}^{N^2-1} \underbrace{ \left( \sum_\alpha  a^\alpha_{k}(t)  {a^\alpha_{l}(t)}^*\right) }_{C_{kl}(t)}
F_k \rho F^\dagger_l =\sum_{k,l=0}^{N^2-1} C_{kl}(t) F_k \rho F^\dagger_l
\end{align}
where the coefficients $C_{kl}(t)$ define a positive definite matrix $C$. The action of $\Lambda_\epsilon$ in the limit $\epsilon \rightarrow 0$ is therefore given by 
\begin{align}
\Lambda_\epsilon \rho &=  C_{00}(\epsilon) \rho + \sum_{k=1}^{N^2-1}  C_{k0}(\epsilon)  F_k \rho +  \sum_{k=1}^{N^2-1}  C^*_{k0}(\epsilon)  \rho  F^\dagger_k + \sum_{k,l=1}^{N^2-1} C_{kl}(\epsilon) F_k \rho F^\dagger_l 
\end{align}
which can be written as
\be
\Lambda_\epsilon \rho = \rho +\epsilon \left( L_0 \rho + \rho L_0^\dagger + \sum_{k,l=1}^{N^2-1} \frac{C_{kl}(\epsilon)}{\epsilon} F_k \rho F^\dagger_l \right)
\label{Linblad01}
\ee
where we have defined
\be
L_0 \equiv \left( \frac{C_{00} (\epsilon) -1 }{2\epsilon} \id + \sum_{k=1}^{N^2-1} \frac{C_{k0}(\epsilon) }{\epsilon} F_k \right)\; . 
\ee
The term in brackets in the expression (\ref{Linblad01}) is therefore identified with the action of the infinitesimal generator $\Li$ of the dynamics by 
$\Lambda_\epsilon \rho = \rho + \epsilon \Li (\rho)$. 
Using the conservation property $\tra{\Lambda_t \rho} =\tra{\rho}$ which is translated on the generator $\Li$ by the condition $\tra{\Li \rho}=0$ we obtain the relation 
\be
L_0+L_0^\dagger = - \sum_{k,l=1}^{N^2-1} \frac{C_{kl}(\epsilon)}{\epsilon} F^\dagger_l F_k \; . 
\ee
The existence of the generator $\Li$ implies that $\lim_{\epsilon\rightarrow 0} \frac{C_{kl}(\epsilon)}{\epsilon} = \gamma^{kl}$ is  independent of the parameter $\epsilon$. 
Decomposing $L_0$  into (\ref{Linblad01})  as a sum of a Hermitian part $(L_0+L^\dagger_0)/2$ plus an anti-Hermitian contribution $-iH/\hslash$, one arrives at the Lindblad equation 
\be
\Li \rho = -\frac{i}{\hslash} [H, \rho]  + \sum_{k,l=1}^{N^2-1} \gamma^{kl} \left(  F_k \rho F^\dagger_l -\frac{1}{2}\left\{ F^\dagger_l  F_k ,\rho \right\} \right)\; . 
\ee
 Since the coefficients $\gamma^{ij}$ define a positive definite Hermitian matrix, we can always put the Lindbladian back into the standard diagonal form \eqref{lindblad2}.

The associated dual Lindblad generator $\Li^*$ is deduced by the duality relation $(A,\Li B)= (\Li^* A, B)$ and it is explicitly given by 
\be
\Li^*(X) =  \frac{i}{\hslash} [H, X] + \sum_{j=1}^{N^2-1}  \left[ L_j^\dagger X L_j - \frac{1}{2} \{ L_j^\dagger L_j, X \}\right] \;, 
\label{linbladdual}
\ee
and one immediately sees that $\Li^* \id =0$.

\subsection{Physical interpretation of the equation}
The time evolution of the system state is governed by the differential equation $\dot{\rho} = {\cal L}(\rho) $ 
which may also be written as 
\be
\dot{\rho} = -  \frac{i}{\hslash} \left( H_{eff} \rho -\rho H_{eff}^\dagger \right) + \sum_{j=1}^{N^2-1}  L_j \rho L^\dagger_j\;,
\ee
where $H_{eff} = i\hslash L_0={H} - i \frac{\hslash}{2}  \sum_{j=1}^{N^2-1}   L^\dagger_j L_j $
defines a so-called non Hermitian Hamiltonian. 
The pseudo-Hamiltonian dynamics generated by  $H_{eff}$  leads to a non-unitary damping process. Indeed,
since the non-unitary part of the effective Hamiltonian $H_{eff}$  is $-i$ times a positive operator, it shifts the spectrum of the Hamiltonian into the left-half complex plane:  $H_{eff} \ket{\psi_n} = (E_n -i \hslash \gamma_n) \ket{\psi_n}$. The partial time evolution of a pure state, generated by $H_{eff}$, is therefore in general a superposition of exponentially damped states  $e^{-i \frac{E_n }{\hslash}t} e^{- \gamma_n t}\ket{\psi_n}$ and leads to a smooth evolution of the state vector. 
Conversely, the so-called jump operators $L_j$, for $j=1,\dots , N^2-1$,  lead to a non-differentiable evolution of the state vector due to the square-root scaling of the associated Kraus operators $K_j\propto  \sqrt{\epsilon} L_j$. Indeed, if we consider the limit $( \ket{\psi (t+\epsilon)} -\ket{\psi(t)})/\epsilon = (\propto \sqrt{\epsilon}L_j(\epsilon) -\id)\ket{\psi(t)}/\epsilon$, as generated by $K_j$, this expression diverges as $\epsilon$ approaches zero. 
Consequently,  the dynamics driven by the Lindblad generator ${\cal L}$ is seen as a combination of a smooth, damped process along with random jumps.

\subsection{Uniqueness of the steady state}
The very form of the Lindblad generator guarantees that it has at least one zero eigenvalue, with associated right eigenvector $\rho^*$ satisfying $\Li \rho^* =0$. 
This steady state is unique if there is no subspace of $\Hs_S$ which is left invariant under the action of the Lindblad generators; see \cite{Evans1977} and \cite{Nigro2019} for a brief review of the problem. More precisely, 
the dynamical semigroup $\Lambda^*_t$, whose infinitesimal generator is ${\cal L}^*$,
is irreducible if and only if the commutant of the set $\{H_S,  L_k, L_k^\dagger\}_{k=1,\dots, N^2-1}$ -- comprising all  jump operators and  the Hamiltonian from \eqref{linbladdual} -- contains only operators proportional to the identity, $\lambda \id$. This implies that under time evolution, the only conserved projector is the identity operator (up to a scalar factor), meaning no proper subspace of $\Hs_S$ remains invariant under the map. As a result, the map is irreducible, and the steady state is unique. In this case, the  Lindblad generator $\cal L$ has a single zero eigenvalue, while all other eigenvalues lie in the open left hand complex plane (i. e., they have  negative real parts). 
The semigroup element $\Lambda_t$ associated to $\cal L$ is  {\it relaxing}, as every initial state  $\rho $ relaxes to $\rho^*$ in the long-time limit.  The uniqueness of the steady state is a necessary and sufficient condition for the semigroup to be relaxing.

\subsection{Collision models}
Collision models, also called repeated interaction models, are aimed at modeling in a very versatile way the contact of a physical system with an environment. 
These models consist of an infinitely large collection of small, independent systems--commonly referred to as ancillae--each interacting with the system one at a time for a finite duration. In order to simulate a Markov map, one usually adds the condition that once an ancilla has interacted with the system, it departs permanently and no longer influences the system in any way. This is reminiscent of Boltzmann's Stosszahlansatz--molecular chaos hypothesis--see \cite{Ciccarello22,Karevski2024} for  comprehensive reviews.  

\begin{figure}[htbp]
   \centering
   \includegraphics[width=7cm]{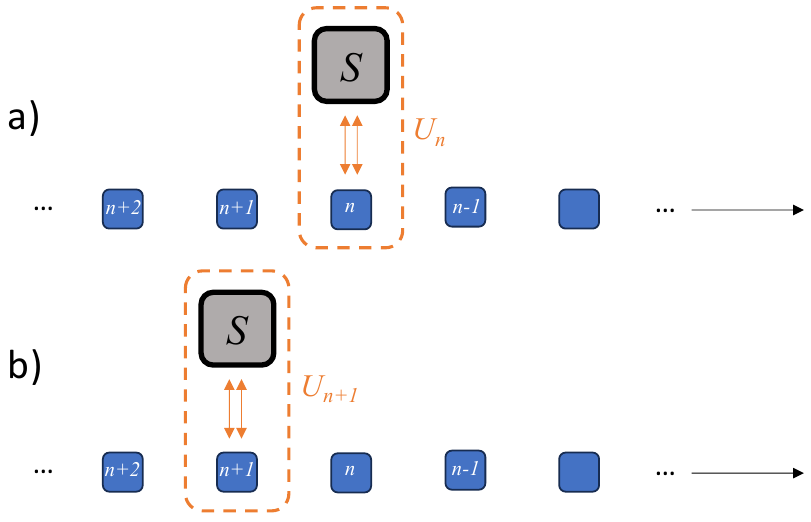} 
     \caption{Schematic representation of the collision map.}
   \label{fig-collision-model}
 \end{figure}

As mentioned above, see figure \ref{fig-collision-model}, in the collision scheme the dynamics takes place through successive interactions of independent ancillae, in general each prepared in the same state, with the system, the most simple situation being that all the successive interactions take the same time $\tau$. In such a scenario, the time evolved state 
of the universe at time $t\in ] (n-1)\tau,n\tau]$, after the $n-1$ first ancilla have interacted with the system, is given by $\omega(t)= U_n(s) \omega((n-1)\tau) U_n^\dagger (s)$ with $U_n(s) = \exp(-i \frac{s}{\hslash} (H_S +H_E +V_n))$. Tracing over the environment, on deduces 
that the system state is given by the recursive equation
 \be
\rho(t) = \tr_{n}  \{ K_n(s) \left[  \, \rho((n-1)\tau)  \otimes \eta_n \, \right] K^\dagger_n(s)  \}\; ,
\ee
where $K_n(s) = \exp(-i \frac{s}{\hslash} (H_S +h_n +V_n))$
is the unitary time evolution operator of the $n^{th}$ ancilla coupled with the system, $\eta_n$ being the state of the $n$th ancilla just before the interaction, $h_n$ being its Hamiltonian  and $V_n$ describing its interaction  with the system. Focusing on the stroboscopic motion, one has for $t=n\tau$
\be
\rho(n\tau) = \tr_{n}  \{ K_n(\tau) \left[ \,  \rho((n-1)\tau) \otimes  \eta_n  \,  \right] K^\dagger_n(\tau)  \} \equiv {\cal K}_\tau[\rho((n-1)\tau) ] \; .
\ee
We see that, if the interactions are always of the same form, the dynamical map ${\cal K}_\tau$ is such that
$\rho \longmapsto {\cal K}_\tau [ \rho] = \tr_{\Hs} \left\{ K(\tau)\, \rho \otimes  \eta \,   K^\dagger(\tau)\right\}$.
The finite evolution of the system state from $\rho_0$, after $n$ collisions can thus be expressed as $\rho(n\tau) = {\cal K}_\tau^n [ \rho_0] $
where ${\cal K}_\tau^n$ stands for the repeated action of ${\cal K}_\tau$, that is $\rho(n\tau) ={\cal K}_\tau [\dots {\cal K}_\tau [ {\cal K}_\tau [ \rho_0] ]\dots ]$. 

Decomposing the unitary operator $K(\tau)$ in a product state basis associated to the system and to the diagonal basis of the ancilla, such that $\eta= \sum_{k=1}^\Omega \pi_k \ket{\phi_k}\bra{\phi_k}$, one can write the collision map ${\cal K}_\tau$ as a  Kraus map ${\cal K}_\tau [\rho] =  \tr_{\Hs} \left\{ K(\tau)\,  \rho \otimes \eta\, K^\dagger(\tau)\right\}  =\sum_{p,q=1}^\Omega  \pi_q   W_p^q(\tau)\, \rho\,  {W_p^q}^\dagger(\tau)$, 
where the generators $W_p^q(\tau) = \bra{\phi_p} K(\tau) \ket{\phi_p}$ are defined on the system Hilbert space. 

To capture the dynamics in the continuous-time limit, the collision duration $\tau$ must tend to zero. However, achieving a meaningful limit requires rescaling the Kraus generators appropriately to ensure that the effects of the interactions remain significant as $\tau\rightarrow 0$. Without this rescaling, the system would effectively decouple from the bath in the limit, resulting in purely unitary dynamics. The non-trivial continuum limit is guaranteed  by expanding $K(\tau)$ as 
\be
K(\tau)=e^{ - i\tau(H_0+V)} = \id -i\tau(H_0+V) -\frac{\tau^2}{2} V^2 + o(\tau)
\ee
where $H_0=H_S+h$ is the free Hamiltonian and $V$ the interaction term. Here, we assume that while $H_0$ is of order one, $V$ scales as ${\cal O}(1/\sqrt{\tau})$ ensuring that $\tau V^2$  remains of order one. This scaling is essential to obtain the non-trivial limit \eqref{linbladdual}. 

The advantage of using collision models is that it is very easy, by specifying the states of the ancillae and the type of interactions with the system, to generate appropriate dissipative channels. In other words, you can build the Lindblad dynamics you want. One can also relax the demand for Markovianity, by allowing ancillae that have already interacted in the past with the system to interact again with it or with future ancillae, thus creating memory effects.

\section{Some concrete situations}
\subsection{Free Fermion models}
Let us consider a free-fermionic system, described by $L_S$ modes and whose unitary free evolution is governed by the free Hamiltonian $H_S= \sum_{i,j=1}^{L_S} T^S_{ij} c_i^\dagger c_j= c^\dagger T^S c$, where $T^S$ is the $L_S\times L_S$ single particle hamiltonian. The $c^\dagger_i$, $c_j$ are the usual fermi operators satisfying the Fermi-Dirac algebra 
$\{ c^\dagger_i, c_j\} = \delta_{ij}$ and $\{ c_i, c_j\} =0$. 
We assume that the system is interacting repeatedly with $L_A$ fermionic ancillae, represented by Fermi-Dirac operators $a^\dagger_i,a_j$ and governed by a free hamiltonian $h = \sum_{i,j}^{L_A} T^A_{ij} a^\dagger_i a_j$, through the bilinear coupling $V= g\sum_{i=1}^{L_S} \sum_{j=1}^{L_A} \Theta_{ij} c^\dagger_i a_j$. 

Starting with an initial Gaussian state, the very form of the bilinear interaction preserves the Gaussianity of the state at all times. A sa consequence, thanks to Wick theorem, the system is completely specified by the two-point matrix $C$ with entries 
\be
C_{ij} (t)= \tr_{S} \{ c^\dagger_j c_i \rho(t)\}\; . 
\ee
It can be shown, see \cite{Karevski2024,Coppola2023b,Coppola2023c,Purkayastha2022}, that in the continuum time limit, the system's dynamics is governed by the Lyapunov equation \cite{Gajic1995}
\be
\dt{}C(t)  = P C(t) + C(t) P^\dagger + F
\label{Lyapunov}
\ee
where the matrix $P =- iT^S - \frac{1}{2} (g\Theta)(g\Theta)^\dagger $ is defined through the single particle hamiltonian $T^S$ and the interaction  matrix $\Theta$
and where $F=  (g \Theta) C^A (g\Theta)^\dagger$ is positive semi-definite and depends on the bath properties through the correlation matrix $C^A$. 
The steady-state solution $C^*$ of the Lyapunov equation is unique for any $F$  if and only if $P$  and $-P^\dagger$ do not share any eigenvalues. 

As a concrete situation, consider a tight binding fermionic chain described by the tridiagonal matrix $T^S$ whose non-zero entries are the first upper and lower diagonals with an homogeneous hopping constant $ T^S_{ii+1}=T^S_{i+1i}= J$. 
Suppose the leftmost(rightmost) site of the chain is in contact with a hot(cold) bath made of the repeated interaction with a single fermion in a state $\eta_{left}$($\eta_{right}$). It has been shown, see \cite{Karevski2024}, that if the baths have identical properties, with the same fermionic populations $n_{left}=n_{right}=n_a$, then the system relaxes toward a steady state with $C^* = n_a \id_s$. That is, the state of the system is completely specified by the local fermionic densities which are specified by the ancilla value $n_a$. 
 On the contrary, when the left and right baths have distinct equilibrium properties defined by their populations $n_{left}$  and $n_{right}$,  a steady particle current flows through the chain over time:  $j= \kappa (n_{left} -n_{right})$ with a constant transport coefficient $\kappa$, showing a ballistic motion of particles \cite{Karevski2009,Platini2010,Karevski2024}. 

\begin{figure}[htbp]
   \centering
   \includegraphics[width=7cm]{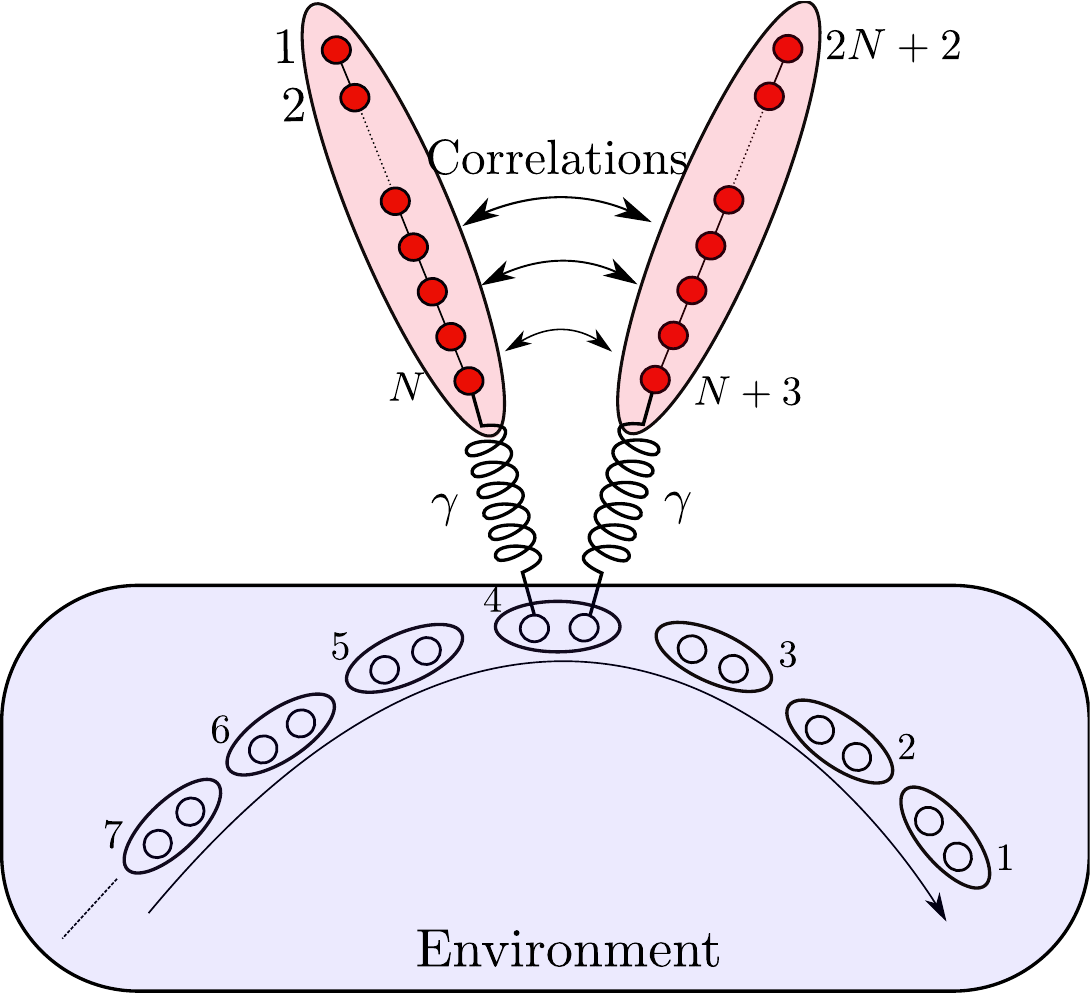} 
     \caption{Schematic representation of the the generation of a rainbow state by the successive interactions with Bell pairs. }
   \label{fig-rep_bell}
 \end{figure}

Another very interesting application concerns the building of entangled pairs which are embodied into material arrays, such as spin or ion chains. 
Entanglement is arguably one of the most counterintuitive phenomena in quantum mechanics, yet it has emerged as a cornerstone of quantum technology, particularly in quantum information processing \cite{Horodecki2009}. Over the past decade, it has been demonstrated that when two quantum chains are locally driven at their boundaries by a shared entangled field, they evolve into a steady state composed of an array of entangled pairs. Each pair consists of one element from the first chain and the other from the second. Remarkably, it has been shown that the field can be fine-tuned to perfectly replicate the entanglement, resulting in a scale-free configuration of two-particle Bell states, known as a rainbow state 
\cite{Zippilli2013,Zippilli2014,Zippilli2015,Zippilli2015b,Wendenbaum2015,Chanda2018,Ma2019,Dutta2020,Dutta2021,Zippilli2021,Pocklington2022,Angeletti2023,Karevski2024}. 
Using the collision model described above with a set of ancillas that are entangled pairs of fermionic modes, it has been shown in \cite{Wendenbaum2015,Karevski2024} that one can replicate the Bell entangled state across two tight-binding chains, see figure \ref{fig-rep_bell}. 

\subsection{Boundary driven Heisenberg spin chains}
As another less trivial example, consider the anisotropic Heisenberg spin 1/2 chain described by the following Hamiltonian
\be
H= J\sum_k \left( \sigma_k^x \sigma_{k+1}^x +  \sigma_k^y \sigma_{k+1}^y +  \Delta \sigma_k^z \sigma_{k+1}^z \right)
\ee
coupled at its left and right boundary spins to two dissipative channels each polarizing the spins in specific directions and described by Lindblad like generators. 
In \cite{Prosen2011,Karevski2013,Popkov2013,Landi2014} it has been shown analytically that the steady state of the Heisenberg chain is given by a Matrix Product state 
involving the $q$-deformed quantum algebra $U_q(SU(2))$, where $q$ is related to the anisotropy parameter $\Delta = (q+q^{-1})/2$. 
The boundary twist leads to non-vanishing stationary currents of all spin components. The matrix product ansatz can be extended to more general quantum systems kept far from equilibrium by Lindblad boundary terms.

\subsection{Atom losses in one-dimensional atomic gas}
\begin{figure}[htbp]
   \centering
   \includegraphics[width=7cm]{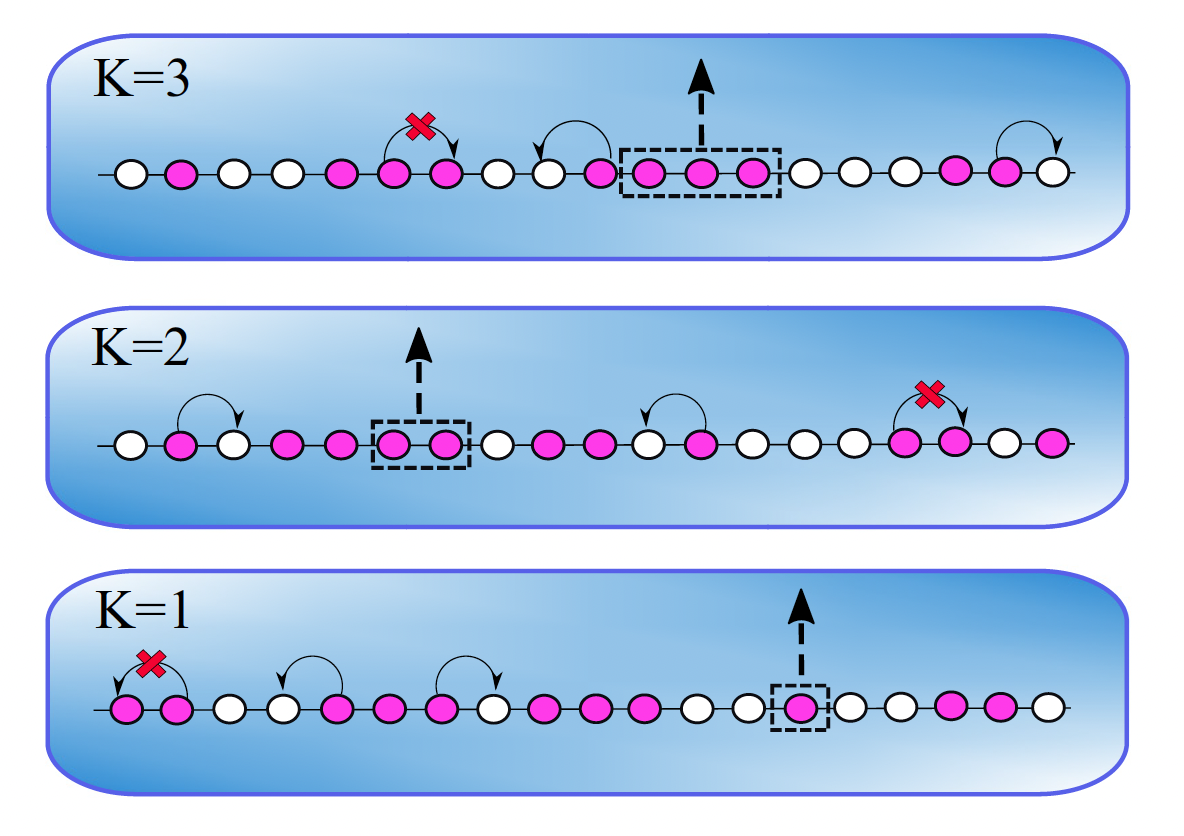} 
     \caption{Schematic representation of the $K$-body losses in a 1d hard-core atomic gas. }
   \label{fig-atom-loss}
 \end{figure}

As a last example, let us illustrate a situation that arises naturally with cold atoms experiments.  Indeed, atom losses are inevitable processes that occur naturally during those experiments. Since this cannot be avoided, it is important to understand their impact on the remaining atoms. A straightforward model to capture this physics involves a lattice gas of hard-core bosons undergoing $K$-body losses (where $K$ represents the number of atoms lost in each event) across neighboring sites, see figure \ref{fig-atom-loss}. The main question is how 
 these losses affect the rapidity distribution $\rho(k)= \moy{c^\dagger(k)c(k)}$ of the atoms. The dynamical equation governing the process is a Lindblad one with jump operators acting at all sites of the lattice gas system and taking the form $L_j = \prod_{l=0}^{K-1} \sigma_{j+l}^- $, where the operator $\sigma_j^-$ annihilate a boson on site $j$. 
 Assuming the losses are sufficiently weak for the system to equilibrate between loss events, it is possible to derive a close expression for the loss functional $F[\rho]$, which characterizes the $K$-body loss process $\dot{\rho}(k) =-F[\rho(k)]$ \cite{Bouchoule2020,Rossini2021, Rosso2022,Riggio2024}.  

For one-body losses, it appears that the loss functional is generally nonlinear and nonlocal in rapidity space, consistent with observations made for the continuous Lieb-Liniger gas in \cite{Bouchoule2020}, and it does not lead the gas to a low-density thermal equilibrium state over time.

For two-body losses, the study of the long-time behavior of both the rapidity distribution and the mean particle density shows that the particle density typically decays as $1/\sqrt{t}$, except when the first Fourier mode of the initial distribution is zero. In this exceptional case, the density decays as $1/t$.  
Inhomogeneous systems consisting in a lattice hardcore bosons gas in the presence of a harmonic potential have also been investigated by numerical methods to solve the
dynamics combining the effects of the losses and of the trapping potential  \cite{Riggio2024}.

\section{Conclusions}
In this brief review, we have presented in detail the general theory of open quantum systems, focusing mainly on the Markovian limit where the dynamics is governed by the so-called Lindblad-Gorini-Kossakowski-Sudarshan equation. 
We have also discussed the main lines of collision models that can be used as a versatile unraveling of the open dynamics. We have shown how these theories can be applied to certain concrete situations involving quantum transport phenomena, the generation and replication of entanglement or even the non-thermal relaxation of cold atomic gases confined in optical traps. 
This article is based on the talk given at the Ohrid 2024 conference. The Society of Physicists of Macedonia and in particular the organisers of the Ohrid conference are warmly thanked.



\end{document}